\documentstyle[epsf,twocolumn]{mn}
\begin{document}

%to produce < and > with a twiddle underneath:
\def\spose#1{\hbox to 0pt{#1\hss}}
\def\ltsima{\mathrel{\spose{\lower 3pt\hbox{$\mathchar"218$}}
     \raise 2.0pt\hbox{$\mathchar"13C$}}}
\def\gtsimafa=0{\mathrel{\spose{\lower 3pt\hbox{$\mathchar"218$}}
     \raise 2.0pt\hbox{$\mathchar"13E$}}}

\title[Effects of Kerr Spacetime on Spectral Features from X-Ray Illuminated
Accretion Discs]{Effects of Kerr Spacetime on Spectral Features from 
X-Ray Illuminated Accretion Discs}

\author[Andrea Martocchia, Vladim{\'\i}r Karas, Giorgio Matt]
{A.\ Martocchia$^1$, V.\ Karas$^2$, G.\ Matt$^3$ \\
$^1$SISSA-ISAS, Via Beirut 2/4, I--34014 Trieste, Italy;
     E-mail: martok@sissa.it \\
$^2$Astronomical Institute, Charles University Prague,
     V Hole\v{s}ovi\v{c}k\'ach 2, CZ--180\,00 Praha, Czech Republic;\\
     E-mail: vladimir.karas@mff.cuni.cz \\
$^3$Dipartimento di Fisica, Universit\`a degli Studi ``Roma Tre'',
     Via della Vasca Navale 84, I--00146 Roma, Italy;\\
     E-mail: matt@haendel.fis.uniroma3.it }

\maketitle

\begin{abstract}

We performed detailed calculations of the relativistic effects acting on
both the reflection continuum and the iron line from accretion discs
around rotating black holes. Fully relativistic transfer of both
illuminating and reprocessed photons has been considered in Kerr
spacetime. We calculated overall spectra, line profiles and integral
quantities, and present their dependences on the black hole angular
momentum. 
We show that the observed EW of the lines is substantially
enlarged when the black hole rotates rapidly and/or the source of
illumination is near above the hole. Therefore, such calculations
provide a way to distinguish among different models of the central 
source.

\end{abstract}


\section{Introduction}

Radiation emitted in the X-ray band by Active Galactic Nuclei (AGN) and
Galactic Black Hole Candidates (BHCs) exhibits the imprints of
strong gravitational fields and orbital rapid motion of matter near a
black hole. Here we adopt the model with a rotating back hole
surrounded by an accretion disc.

The very first studies of light propagation in the Kerr metric were
performed in the 1970's (Bardeen, Press \& Teukolsky 1972; Cunningham \&
Bardeen 1973; Cunningham 1975), and it was argued that the observed
radiation should be substantially affected by the presence of a black hole 
and by its rotation. In the last few years, even before the great excitement
aroused by the ASCA detection of the relativistic iron K$\alpha$ line in
the spectrum of the Seyfert 1 galaxy MCG-6-30-15 (Tanaka et al. 1995),
many authors modelled the effects of Special and General Relativity on
the line profiles under various physical and geometrical assumptions (e.g.
Fabian et al. 1989).
Calculations of line profiles from a disc-like source in Kerr spacetime
have been performed by Laor (1991), Kojima (1991), Hameury, Marck \&
Pelat (1994), Karas, Lanza \& Vokrouhlick\'y (1995), Bromley, Chen \&
Miller (1997), Fanton et al.\ (1997), Dabrowski et al. (1997),
\v{C}ade\v{z}, Fanton \& Calvani (1998), and others. Nowadays, these
studies are particularly relevant for the understanding of
AGN in view of the above mentioned detection of line profiles which
suggests substantial relativistic effects in many objects (Nandra et al.\
1997).

Various simplifying assumptions have been adopted in previous
calculations. Line profiles have been often calculated independently of
the underlying reflection continuum (Guilbert \& Rees 1988; Lightman \&
White 1988), which is however produced along with the line after
illumination of the disc by some primary X-ray source. Even when
considered, calculations of light propagation were performed in
Schwarzschild metric. Matt, Perola \& Piro (1991) adopted a weak-field
approximation. Recently, Macio{\l}ek-Nied{\'z}wiecki \& Magdziardz
(1998), and Bao, Wiita \& Hadrava (1998) made use of fully relativistic
codes, but still in Schwarzschild metric. Morever, a simple power-law
parameterization of the disc emissivity has been usually adopted, e.g.\
the one which follows from the Page \& Thorne (1974) model, while the
actual emissivity is substantially more complex and depends on the
geometry of the illuminating matter (Matt, Perola \& Piro 1991;
Martocchia \& Matt 1996).

Self-consistent calculations of iron lines and continuum together are
still missing in the case of Kerr metric. This problem is thus examined
in the present paper. Reflected light rays are properly treated as
geodesics in the curved spacetime and, furthermore, light propagation
from the primary, illuminating source to the reflecting material is
also calculated in a fully relativistic approach. The adopted point-like
geometry for the primary X-ray emitting region (Sec.\ 2.1) is clearly a
simplification, but can be considered as a rough phenomenological
approximation to more realistic models. 
Off-axis flares, which would be expected from magnetic reconnection above
the accretion disc, have been considered by Yu \& Lu (1999) and Reynolds
et al.\ (1999) in Schwarzschild and Kerr metric, respectively. More complex
scenarios, like hot coronae and non-keplerian accretion flows, would be
very interesting to explore but are beyond the scope of this paper.

\section{An illuminated disc in Kerr metric}

\begin{figure}
\epsfxsize=\hsize
\epsffile{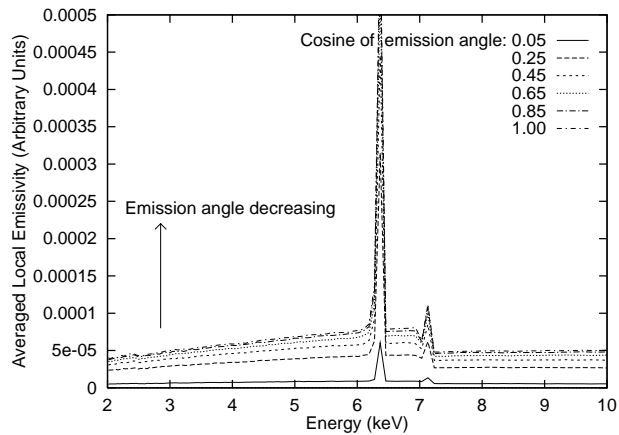}
\caption{ Examples of local spectra $F(E)$ around the iron line
rest energy, $E_0 = 6.4$ keV. Only angular dependences of the outgoing 
flux are shown here: $F$ represents the radiation flux averaged over the
incident angle of the photons from the primary source.}
\label{fig_avera}
\end{figure}

\subsection{The model}

We adopt the model described by Martocchia \& Matt (1996): a
geometrically thin equatorial disc of cold (neutral) matter, 
illuminated by a stationary point-like source on the symmetry axis at
height $h$. The local emissivity of the disc has been computed following
Matt, Perola \& Piro (1991), taking into account the energy and
impinging angle of illuminating photons, as seen by the rotating matter
in the disc. Then, transfer of photons leaving the disc was carried out
according to Karas, Vokrouhlick\'y \& Polnarev (1992).

A similar model of illumination has been used by Henry \& Petrucci
(1997), who call it an {\em{anisotropic illumination model\/}}, 
and by other authors (Reynolds et al.\ 1999, and references cited
therein). Bao, Wiita \& Hadrava (1998) used a fixed direction of
impinging photons (as if the source were distant, possibly displaced
from the rotation axis); however, these authors did not solve the
radiation transfer within the disc. This is an important point, as
different values of $h$ correspond to substantially different
illumination of the disc in the local frame corotating with the matter,
and consequently to different emissivity laws, $I(r,h)$. With decreasing
$h$, the effect of light bending is enhanced and the fraction of X-ray
photons impinging onto the disc is increased with respect to those
escaping to infinity and contributing to the direct (primary) continuum
component. Moreover, photons arriving to the disc at radii $\ltsima{h}$
are blueshifted, so that the fraction of photons with energies above the
photoionization threshold is increased.

It has been argued (e.g.\ Martocchia \& Matt 1996) that a way to
discriminate between static and spinning black holes could be based on
the fact that the innermost stable orbit $r_{\rm{ms}}(a,m)$ of a
fast-rotating black hole lies close to the event horizon and approaches
the gravitational radius $r_{\rm{g}}(a,m)\rightarrow{m}$ for a maximally
rotating Kerr hole with the limiting value of $a\rightarrow{m}$ 
(we use standard notation,
Boyer-Lindquist coordinates\footnote{We recall that the Boyer-Lindquist
radial coordinate is directly related to the circumference of
$r={\rm{const}}$ circles in the equatorial plane. Coordinate separation
between $r_{\rm{ms}}(a,m)$ and $r_{\rm{g}}$ obviously decreases when
$a\rightarrow{m}$, but proper radial distance (which has direct physical
meaning) between these two circles increases as $(m-a)^{-1/6}$. What is
however essential for our discussion of observed radiation fluxes are
local emissivities and the total outgoing flux, which is obtained by
summing over individual contributions of $r={\rm{const}}$ rings. When
expressed in terms of $r$, as we see in Fig.~\ref{fig_emis}, the local
emissivity is large near the inner edge and becomes very anisotropic
when $a$ approaches its limiting value for the maximally rotating hole.}
and geometrized units $c=G=1$; e.g. Chandrasekhar 1983).

Highly redshifted features would then represent an imprint of photons
emitted at extremely small disc radii, which is possible only near
fast-rotating black holes. Other explanations are also viable but
require more complicated models of the source (compared to purely
Keplerian, geometrically thin discs). Reynolds \& Begelman (1997)
pointed out that the difference between spectra of rapidly versus slowly
rotating black holes would be much smaller if efficient line emission is
allowed also from free-falling matter inside the last stable orbit, and
they applied this assumption to the reddish line profile observed during
a low-flux state of MCG--6-30-15 (Iwasawa et al.\ 1996). If this is the
case, the presence of an extended red tail of the line could no longer
be used as an evidence for rapid rotation of the black hole, whereas
validity of the ``spin paradigm'' (the often made suggestion that
rotating black holes are associated with jet production and
radio-loudness) remains preserved. The problem can be solved by
calculating in detail the optical thickness and the ionization state of
the free-falling matter, as in Young, Ross \& Fabian (1998), who noted
that the reflection component in MCG--6-30-15 is not consistent with the
expected ionization state of the matter inside $r_{\rm{ms}}$.

\subsection{The emissivity laws}

\begin{figure}
\epsfxsize=\hsize
\epsffile{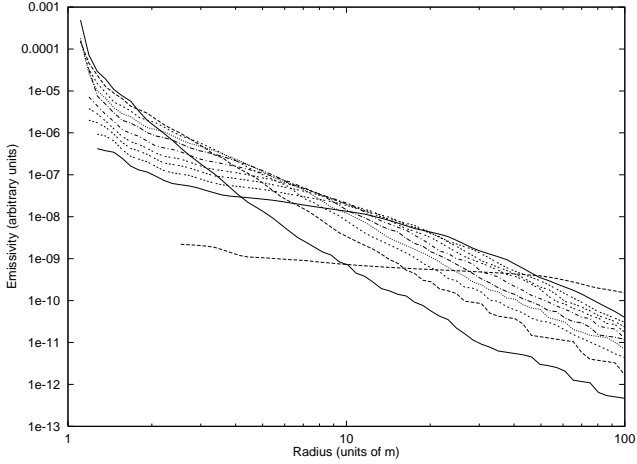}
\caption{ Emissivity laws $I(r)$ (in arbitrary units) corresponding to
different source heights above an extreme Kerr hole, as discussed in
Martocchia \& Matt (1997). Looking at the
right side of the diagram the curves correspond, from top to bottom,
to heights of 100, 20, 15, 12, 10, 8, 6, 5, 4, 3, 2 (units of $m$). 
The curves steepen when $h$ decreases, which corresponds to
increasing anisotropy of emission. }
\label{fig_emis}
\end{figure}

We used a Monte Carlo code to calculate photon transfer within the
disc (Matt, Perola \& Piro 1991). The resulting local spectra in the
frame comoving with the disc matter are shown in Figure \ref{fig_avera}.
Assumptions about local emissivity, disc shape and rotation law can be
varied in our code in order to account for different accretion models,
but here we describe only the case of standard Keplerian, geometrically
thin and optically thick discs for simplicity. Let us note that any
radial inflow decreases the observed line widths when compared with the
corresponding case of a Keplerian disc. While in other works the disc
emissivity has been often described as a power law ($\propto{r^{-s}}$), we
made use of the emissivities derived by Martocchia \& Matt (1997)
through integration of geodesics from the primary source
(Fig.~\ref{fig_emis}). The source distance $h$ stays here, instead of
the power law index $s$, as one of the model parameters.

\subsection{Spectral features}

\begin{figure}
\epsfxsize=\hsize
\epsffile{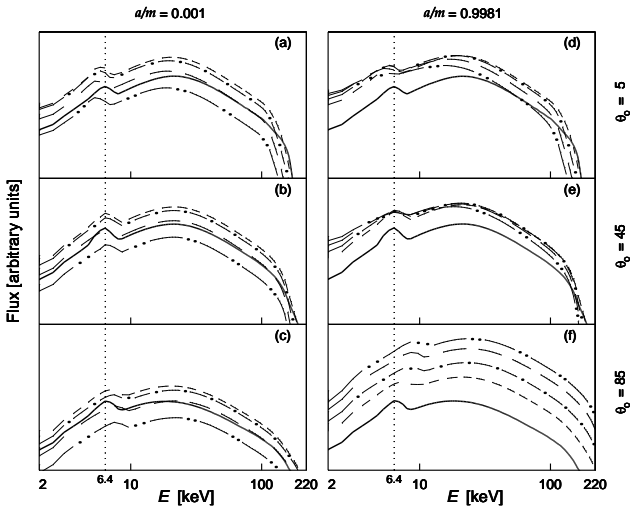}
\caption{ Examples of broad-band (2--220 keV) spectra for three 
inclination angles ($\theta_{\rm{o}}=5$, $45$, and $85$ degrees). The
following types of lines have been used in this graph to indicate
the source height (in Boyer-Lindquist coordinates):
$h=100m$ --- thick solid line; 
$h=20m$ --- short-dashed;
$h=10m$ --- dot-dashed; 
$h=6m$ --- long-dashed; 
$h=4m$ --- double-dot-dashed.
Left panel corresponds to a slowly rotating black hole,
while right panel shows the case of rapid rotation, as indicated
on the top. Here we considered a disc
extending down to the innermost stable orbit. Outer edge is at
$r_{\rm{out}} = 100m$. }
\label{fig_bro100}
\end{figure}

\begin{figure}
\epsfxsize=\hsize
\epsffile{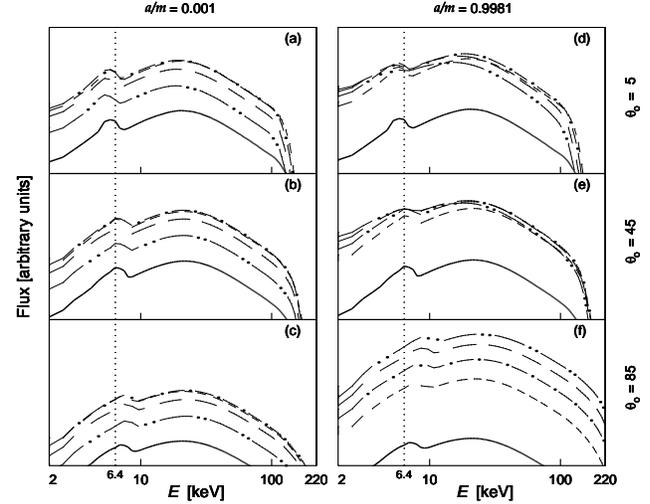}
\caption{ As in the previous figure, but for
{\bf $r_{\rm{out}} = 34m$}. }
\label{fig_bro34}
\end{figure}

Illumination of cold matter in the disc by the primary, hard X-ray flux
results in a Compton-reflection component with specific signatures of
bound-free absorption and fluorescence. The most prominent iron features
are gathered in a narrow energy range: K$\alpha$ and K$\beta$ lines with
rest energies at 6.4 keV and 7.07 keV, respectively, and the iron edge
at 7.1 keV. On the other hand, the overall continuum is rather broad,
and it is best illustrated in the energy range $E=2$--$220$ keV (Figures
\ref{fig_bro100} and~\ref{fig_bro34}). The continuum gets broader with
increasing inclination due to Doppler shifts. The large spread in blue-
and red-shifts blurs the photoelectric edge at 7.1 keV and results 
in broad troughs. This can be seen better after increasing the
energy resolution, i.e.\ in the narrow band spectra (next section). The
overall spectrum is also sensitive to inclination angles, being more
intense when seen pole-on ($\theta_{\rm{o}}=0$).

Line profiles in real spectra must result from a subtraction of the
proper underlying continuum, taking into account the relativistic
smearing of the iron edge. This work has been already started by several
authors in Schwarzschild metric (Macio{\l}ek-Nied{\'z}wiecki \&
Magdziarz 1998; Young, Ross \& Fabian 1998; $\dot{\rm Z}$ycki, Done \&
Smith 1997 and 1998) and developed further here, in the case of sources
around rotating black holes. The effects of inclination on the smearing
and smoothing of all spectral features may be dramatic in Kerr metric
not only because of substantial energy shifts of photons emitted at the
innermost radii, but also due to the mutual combination of this effect on
both the iron line and the underlying continuum. The spectral features
thus spread across a broad range of energies and may become difficult to
observe. Similar behaviour could be obtained for static black holes if
efficient emission were allowed from inside the innermost stable orbit,
but Young, Ross \& Fabian (1998) noticed that in this case a large
absorption edge beyond 6.9 keV should appear, and this is usually not
observed in AGN.

\subsection{The line profile}

The qualitative behaviour of the observed line profiles as a function of 
observer's inclination is very intuitive.
When a disc is observed almost pole-on, the iron line gets somewhat
broadened and redshifted because of the deep potential well and transverse
Doppler effect due to the rapid
orbital motion of the matter. This effect is more
pronounced in the extreme Kerr case, when the emitting material, still
on stable orbits, extends down almost to the very horizon. The
broadening of the observed spectral features is particularly evident
when strongly anisotropic emissivity laws, resulting from small $h$, are
considered. As the disc inclination increases, the iron line becomes
substantially broader, with 
the well-known double-peaked profile due to the Doppler shift of the
radiation coming from opposite sides of the disc. The interplay of
Doppler effects and gravitational light bending determines the details of
the profile. The relative distance of the two horns increases with the
inclination angle; then the horns and the iron edge almost disappear as
individual and well recognizable features for very high inclinations
when the Doppler effect is maximum (disc observed edge-on). Therefore,
such horns are well visible only at intermediate inclinations. In this
situation the blue peak is substantially higher than the red one, due to
Doppler boosting. Quantitative account of all these effects requires to
adopt specific models and to calculate profiles numerically.

\begin{figure}
\epsfxsize=\hsize
\epsffile{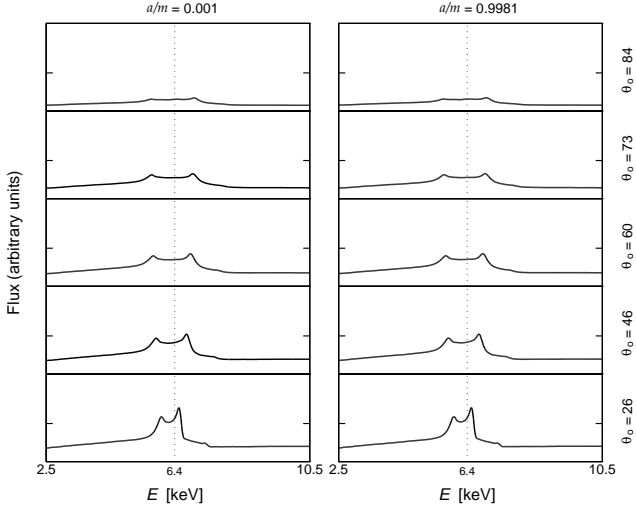}
\caption{ Narrow-band (2.5--10.5 keV) spectra. Calculations have been
performed for both non-rotating (left) and maximally rotating (right)
holes. The emitting area is comprised between the innermost stable orbit
($r_{\rm{in}}=r_{\rm{ms}}=6m$ and $1.23m$, respectively) and
$r_{\rm{out}} = 100m$. Inclination angles are (from top to
bottom): $\cos{\theta_{\rm{o}}}=$ 0.1, 0.3, 0.5, 0.7, and 0.9. 
$h=100m$. The vertical axis is in logarithmic scale. }
\label{fig_h100}
\end{figure}

\begin{figure}
\epsfxsize=\hsize
\epsffile{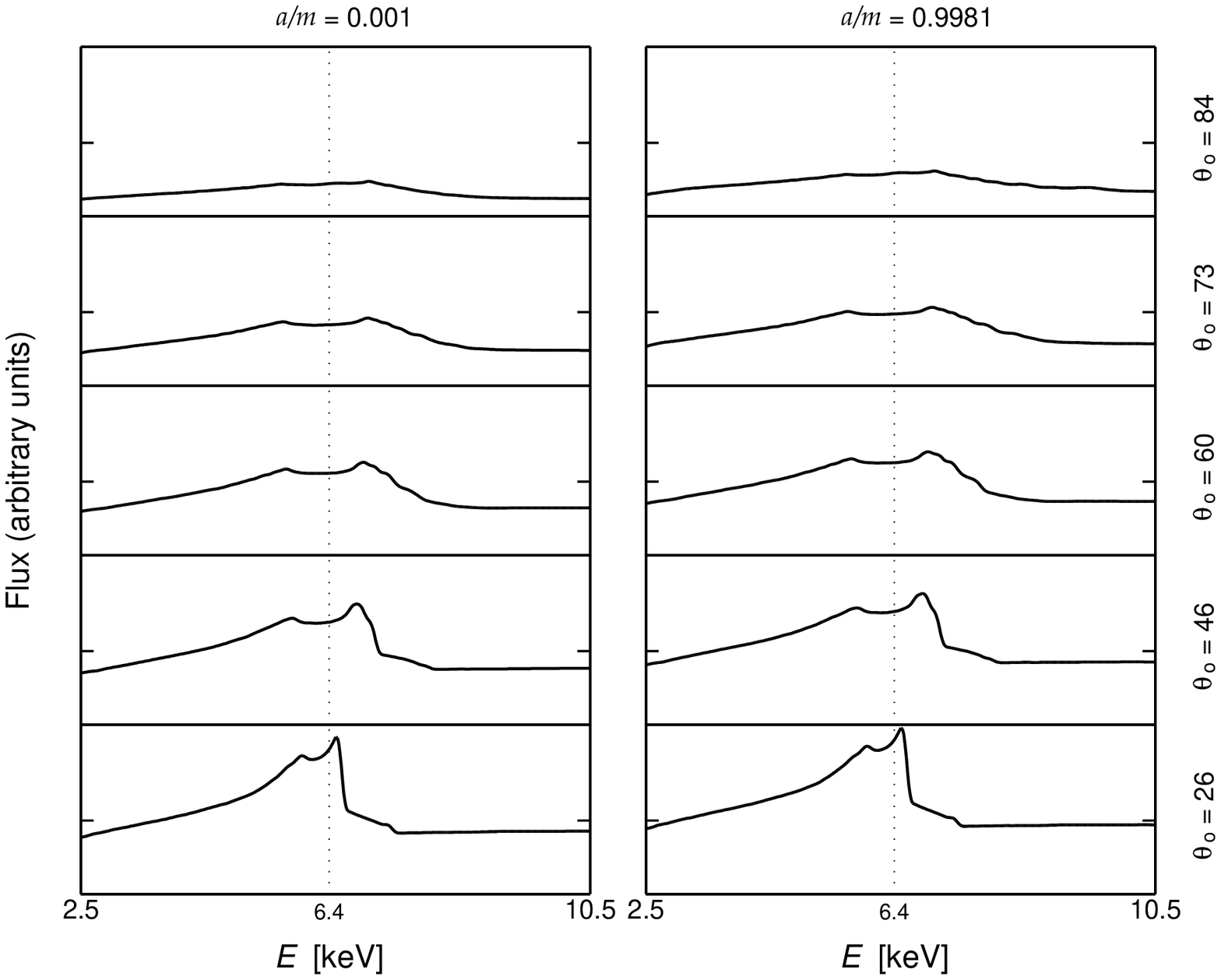}
\caption{ As in previous figure but for $h=20m$. }
\label{fig_h20}
\end{figure}

\begin{figure}
\epsfxsize=\hsize
\epsffile{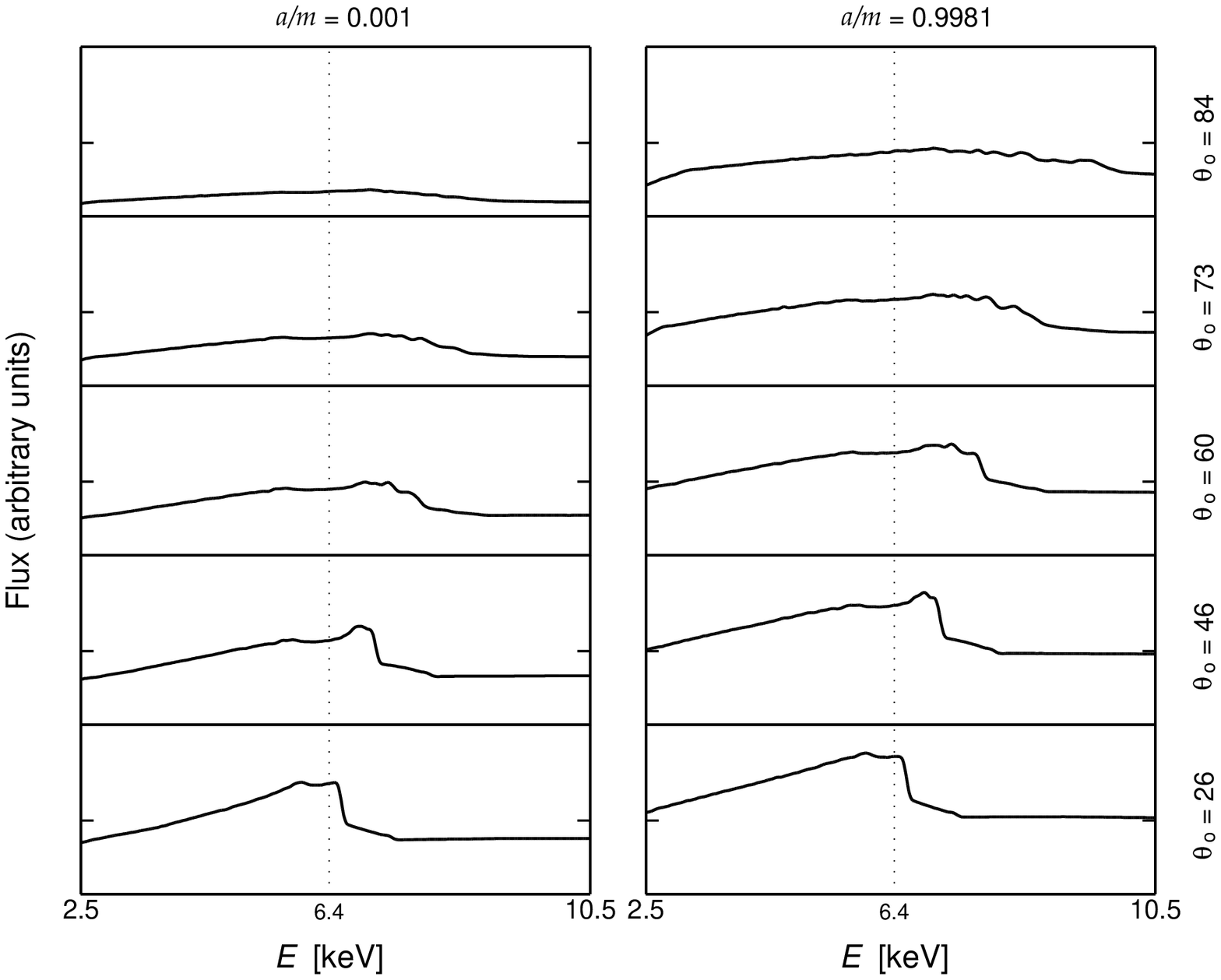}
\caption{ As in previous figure but for $h=10m$. }
\label{fig_h10}
\end{figure}

\begin{figure}
\epsfxsize=\hsize
\epsffile{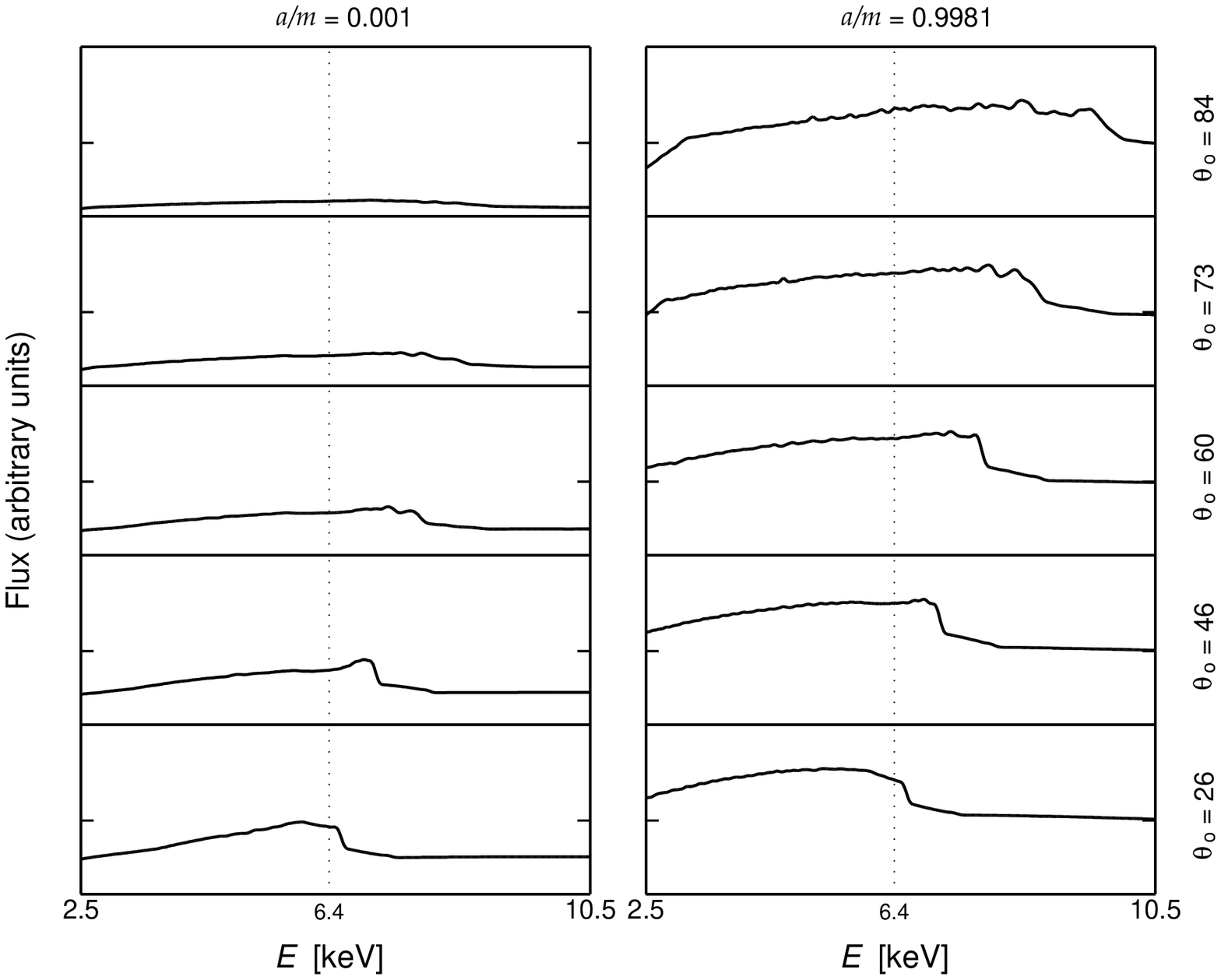}
\caption{ As in previous figure but for $h=6m$. }
\label{fig_h6}
\end{figure}

\begin{figure}
\epsfxsize=\hsize
\epsffile{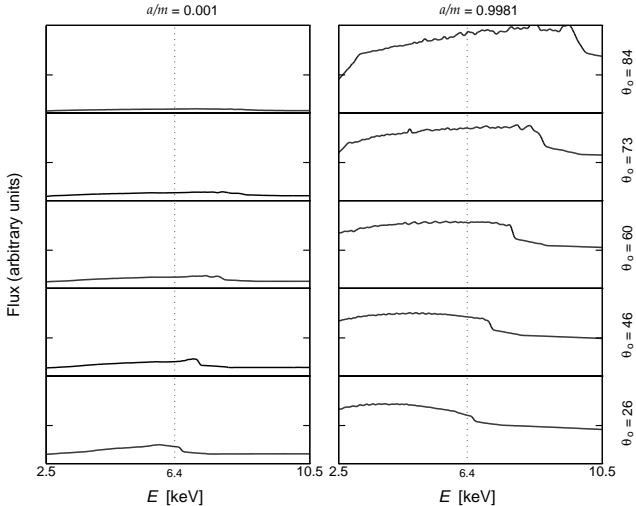}
\caption{ As in previous figure but for $h=4m$. }
\label{fig_h4}
\end{figure}

\begin{figure*}
\epsfxsize=\hsize
\epsffile{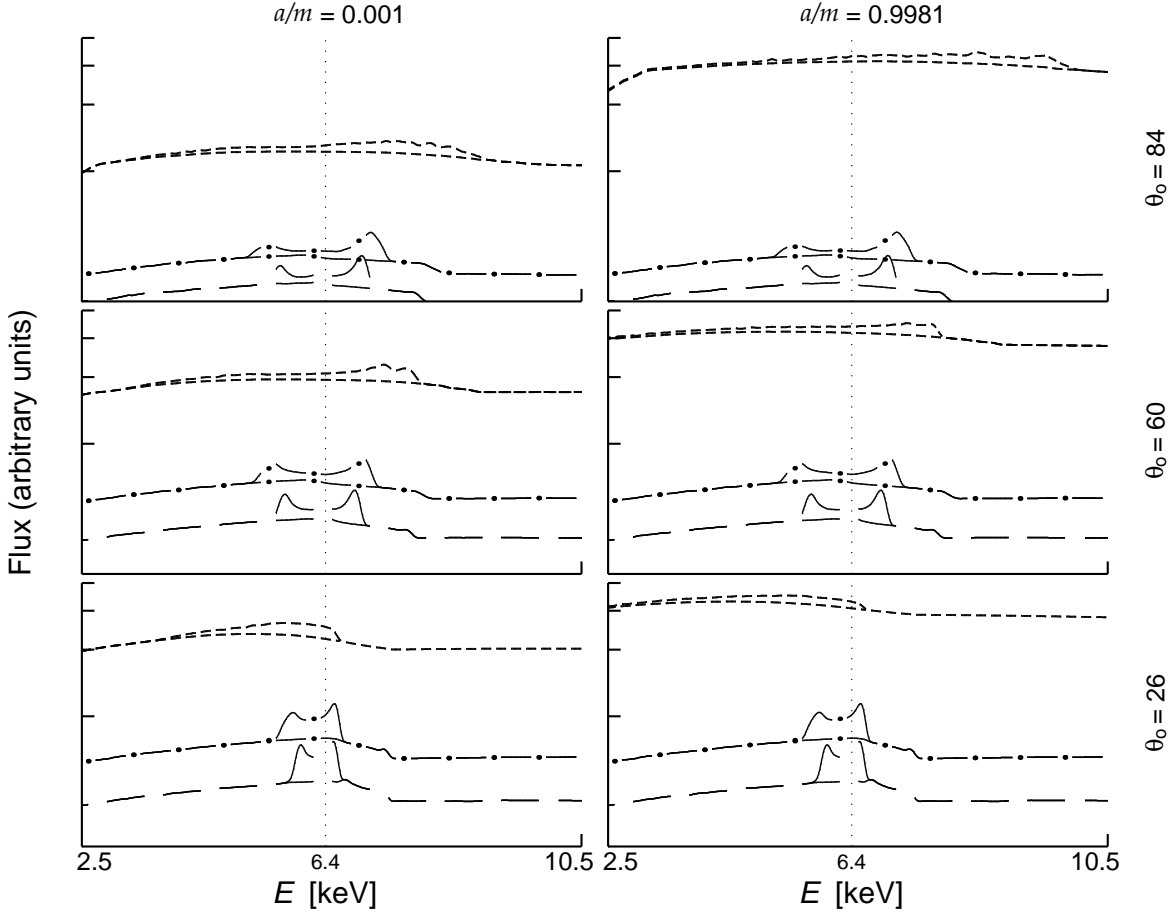}
\caption{Contributions to the total profile for $h=6m$ from three
subsections of the disc are shown in this figure: $r/m$ extending from
$r_{\rm{ms}}$ up to
34 (short-dashed), 34--67 (dash-dotted), and 67--100 (long-dashed).
Underlying curves of the reflection continuum are also plotted. Three
values of inclinations (corresponding to $\cos{\theta_{\rm{o}}}=0.1$,
$0.5$, $0.9$) are considered. Vertical scale is logarithmic.}
\label{fig_h6all}
\end{figure*}

Figures~\ref{fig_h100}--\ref{fig_h6all} show the line profiles 
corresponding to different $h$ in our model. It is evident that the effects 
of the anisotropic illumination can be enormous, causing a substantial
amount of the re-emitted flux to be highly redshifted, especially in the
low-$h$ case. When the source is very close to the black hole,
because of strong anisotropy only the innermost part of the disc
contributes to the line and to the reflection continuum fluxes. As a
consequence, spectral features can be huge in Kerr metric, whereas in
the Schwarzschild case they gradually disappear if no efficient
reemission is possible from $r<6m$.

The adopted emissivity law is clearly a key ingredient in the
calculations of reflected spectra. Flat local emissivity laws apply
when the source is distant from the hole ($h{\gg}r_{\rm{g}}$), and
result in spectra which show very weak dependence on the black hole
angular momentum. On the other hand, with steep emissivities (low
$h$) the observed spectra strongly depend on $a/m$.

The detailed line profiles contain a wealth of information which can in
principle be compared with observed data, but it may be useful to
describe them also by integral quantities which can be determined also
from data with lower resolution. In the next section
we will consider the line equivalent width (EW), the centroid energy
($E_{\rm{cen}}$), and the geometrical width ($\sigma$). Here, EW is
defined in terms of radiation fluxes (line and continuum) as
${\rm{EW}}=\int{F_{\rm{line}}(E){\rm{d}}E}/
F_{\rm{cont}}(E=E_{\rm{line}})$, where the underlying continuum can be
either the direct, or the reflected one, or their sum; i.e.\
$F_{\rm{cont}}=F_{\rm{dir}}+F_{\rm{ref}}$. $E_{\rm{line}}$ is the rest
energy of the line, i.e. 6.4 keV  for the iron K$\alpha$ line. 

\section{Combined effects and integral quantities}
\label{effects}

\begin{figure}
\epsfxsize=\hsize
\epsffile{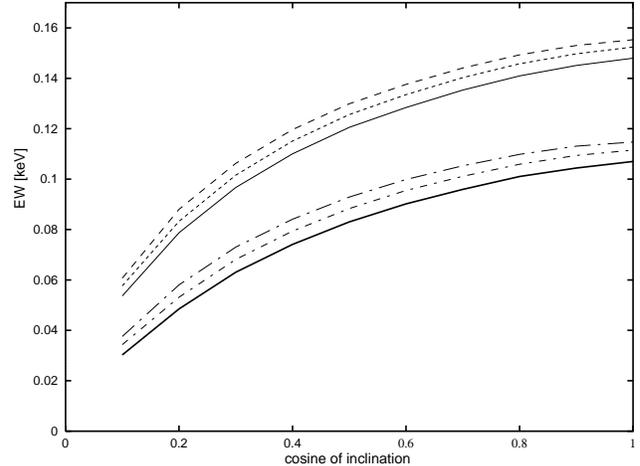}
\caption{ Equivalent width as a function of the cosine of 
inclination. Here, $h = 20m$, $a/m=0.9981$.
For such a large value of $h$, even with almost extreme value of 
$a/m$, the results are very similar to the non-rotating case
and thus agree well with the values in Matt et al.\ (1992). Here EWs refer 
to the direct continuum only, and the solid angle distortion is negligible
($f \sim 1$). See the text for more details. }
\label{fig_ew100e1000}
\end{figure}

\begin{figure}
\epsfxsize=\hsize
\epsffile{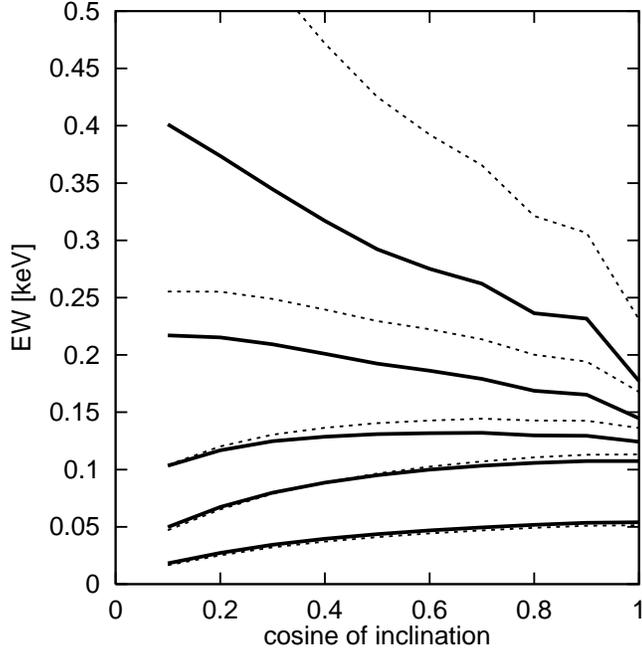}
\caption{ Equivalent width versus cosine of the observer's inclination
angle, for different illuminations and a disc extending from the innermost
stable orbit up to $100m$. The solid lines refer to EWs with respect to
the total underlying continuum, the dashed ones to EWs with respect to the
direct contribution only. From top to bottom, the curves are for a
source located at $h/m=4$, $6$, $10$, $20$, and $100$, respectively. The
black hole spin parameter is $a/m = 0.9981$.  }
\label{fig_h_inc}
\end{figure}

%\begin{figure}
%\epsfxsize=\hsize
%\epsffile{ew_34.eps}
%\caption{ As in the previous figure (Fig.12) but for
%the innermost part of the disc, up to $r=34m$ only.
%It is clear that the EW values are substantially different than in the 
%$r_{out}=100m$ case only
%when the source is located far away from the hole ($h=20m$, $100m$) because
%in this case much of the primary flux gets lost. }
%\label{fig_ew_34}
%\end{figure}
%\begin{figure}
%\epsfxsize=\hsize
%\epsffile{ew_h34.eps}
%\caption{ As in previous figure, but for $r_{\rm{out}} = 34m$. In this
%case the EW decreases more rapidly with source height because the solid
%angle of primary flux impinging onto the disc becomes lower when $h$
%increases. Like before, the three curves correspond to, from top
%to bottom: $a=0.9981m$, $a=0.5m$ and $a=0.001m$.}
%\label{fig_ew_h34}
%\end{figure}

\begin{figure}
\epsfxsize=\hsize
\epsffile{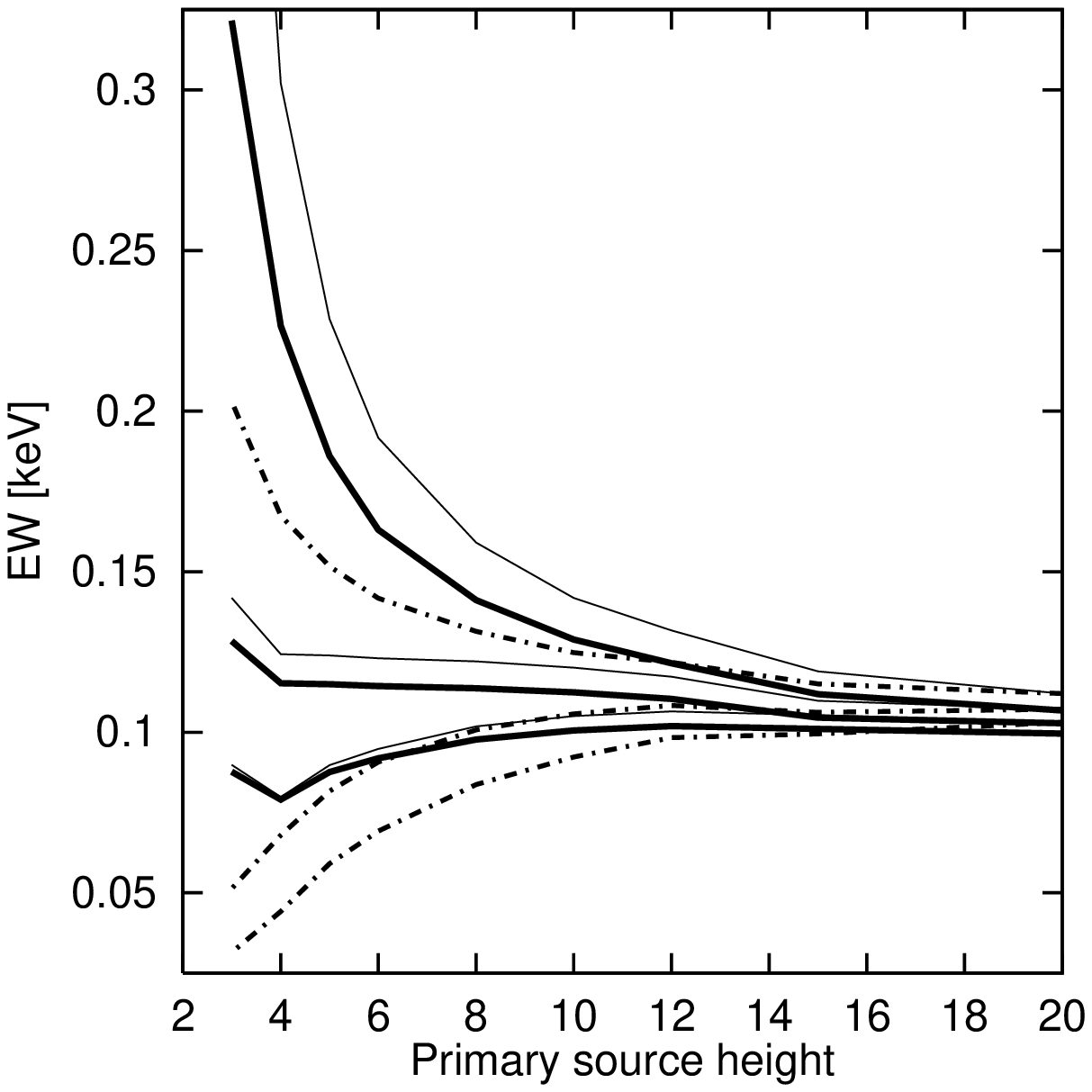}
\caption{ EW dependence on the source height $h$, for an observer
inclination fixed at 30 degrees and a disc extending from the innermost
stable orbit up to $100m$. Three values have been considered of the
black hole spin (from top to bottom): $a = 0.9981m$, $0.5m$ and
$0.001m$. See the text for more details. }
\label{fig_30degree}
\end{figure}

When dealing with low energy resolution detectors 
and/or faint sources, a detailed
study of the line profile may not be possible. In these cases, one can
resort to integral quantities. In
Figs.~\ref{fig_ew100e1000}--\ref{fig_30degree} we present the equivalent
widths of the iron K$\alpha$ 
fluorescent line in different cases. As a check, we verified 
that in the Schwarzschild case the results were in agreement with those
of Matt et al. (1992).

We carried out calculations for several different values of the model
parameters $h$, $a$, $\theta_{\rm{o}}$, and for different sizes of the
disc. Apart from the above mentioned strong $h$-dependencies of the observed
spectra, it turns out that the radial extension of the disc is an important
parameter determining the reflection continuum and affecting EWs. 

In Fig.~\ref{fig_ew100e1000}, the height of the source is fixed to $20m$.
The upper three curves correspond to an outer radius
$r_{\rm{out}}=1000m$, while those below to $r_{\rm{out}}=100m$. For both sets,
the curves refer to (from top to bottom):
(i)~$r>r_{\rm{in}}=1.23m$ (the innermost stable orbit in Kerr metric,
given the fiducial value of $a/m=0.9981$, Thorne 1974);
(ii)~$r>r_{\rm{in}}=6m$ (the innermost stable orbit in Schwarzschild
metric): we found that the resulting values for static and for rotating
holes are very similar in this case; (iii)~$r>r_{\rm{in}}=10m$:
values for static and for rotating black holes are almost identical.

%-----------------------------
\begin{table*}
\centering
\label{tab_1}
%\vspace{0.5pt}
\begin{tabular}{cccccccccccc}
\hline
\hline
\multicolumn{2}{l}{$\cos{\theta_{\rm{o}}}$}~~ &
\multicolumn{2}{c}{$h = 4m$} &
\multicolumn{2}{c}{$h = 6m$} &
\multicolumn{2}{c}{$h = 10m$}&
\multicolumn{2}{c}{$h = 20m$} &
\multicolumn{2}{c}{$h = 100m$} \\
\hline
   ~     & $E_{\rm{cen}}$           &  7.65 &  (6.77) & 7.45 &  (6.73)  &
7.18 &  (6.64) & 6.78 &  (6.52) & 6.44 &  (6.44) \\
  0.1    & $\sigma$                 &  1.71 &  (1.29) & 1.71 &  (1.27)  &
1.62 &  (1.17) & 1.31 &  (0.96) & 0.64 &  (0.61) \\
   ~     & EW                       &  401  &  (30)   & 217  &  (35)    &
103  &  (38)   & 50   &  (36)   & 18   &  (18) \\
\hline
   ~     & $E_{\rm{cen}}$           &  6.50 &  (6.65) & 6.56 &  (6.63)  &
6.56 &  (6.57) & 6.49 &  (6.47) & 6.38 &  (6.38) \\
  0.3    & $\sigma$                 &  1.63 &  (1.26) & 1.56 &  (1.23)  &
1.38 &  (1.14) & 1.07 &  (0.96) & 0.64 &  (0.62) \\
   ~     & EW                       &  345  &  (58)   & 209  &  (66)    &
125  &  (72)   & 80   &  (68)   & 34   &  (34) \\
\hline
   ~     & $E_{\rm{cen}}$           &  5.71 &  (6.36) & 5.97 &  (6.36)  &
6.17 &  (6.35) & 6.29 &  (6.34) & 6.34 &  (6.34) \\
  0.5    & $\sigma$                 &  1.50 &  (1.13) & 1.42 &  (1.10)  &
1.23 &  (1.02) & 0.95 &  (0.86) & 0.58 &  (0.57) \\
   ~     & EW                       &  292  &  (72)   & 192  &  (82)    &
131  &  (89)   & 95   &  (85)   & 44   &  (43) \\
\hline
   ~     & $E_{\rm{cen}}$           &  5.08 &  (6.01) & 5.47 &  (6.03)  &
5.82 &  (6.08) & 6.09 &  (6.17) & 6.27 &  (6.27) \\
  0.7    & $\sigma$                 &  1.35 &  (0.93) & 1.26 &  (0.91)  &
1.08 &  (0.84) & 0.82 &  (0.71) & 0.48 &  (0.47) \\
   ~     & EW                       &  262  &  (77)   & 179  &  (88)    &
132  &  (97)   & 103  &  (95)   & 49   &  (49) \\
\hline
   ~     & $E_{\rm{cen}}$           &  4.53 &  (5.64) & 5.01 &  (5.68)  &
5.49 &  (5.79) & 5.88 &  (5.97) & 6.19 &  (6.19) \\
  0.9    & $\sigma$                 &  1.20 &  (0.67) & 1.11 &  (0.65)  &
0.92 &  (0.61) & 0.66 &  (0.51) & 0.33 &  (0.31) \\
   ~     & EW                       &  232  &  (79)   & 165  &  (91)    &
130  &  (101)  & 107  &  (100)  & 53   &  (53) \\
\hline  
\hline
\end{tabular}
\caption{ Table of integral quantities characterizing the iron K$\alpha$
line profiles:
centroid energy and  geometrical width (in eV), and equivalent width
(in keV). Comparison
is given between corresponding quantities for a fast rotating and
a non-rotating (in parenthesis) black holes.}
%\label{table} \vspace(0.5pt)
\end{table*}

In the other figures the strong dependence on the source parameter $h$
is evident. In our calculations we also accounted for the effect of the
light-rays distortion on the primary flux, i.e. for the fact that the
solid angle $\Omega$ of the direct continuum which escapes to infinity 
(arriving directly from the primary source to the observer) diminishes
considerably when the source is near the black hole (low $h$). The effect is
accounted for by introducing an $h$-dependent factor 
$f(h)=\Omega_{\inf}/\Omega_{\inf, {\rm cl}}$ which
multiplies $F_{\rm{dir}}$ in the definition of EW. As expected, 
$f$ approaches unity if the source is far away from the hole.

Figure~\ref{fig_h_inc} shows the equivalent width as a function of
the cosine of inclination, $\mu=\cos\theta_{\rm{o}}$. The slope of 
EW$(\mu)$ gets inverted around $h\sim10m$: for larger values of $h$ we 
obtain EW {\em{decreasing}\/} with $\mu$, as in the Schwarzschild case,
whereas for $h\ltsima10m$ EW {\em{increases}\/} when $\mu$ decreases.
This is a direct consequence of both the large efficiency of
line emission and the enhanced influence of light bending
from the innermost Kerr orbits, which strongly affects the
profiles. This behaviour is less pronounced when EWs with respect
to the total continuum (direct plus reflected; cf.\ solid lines) are
considered, the reason being that the Compton-reflected contribution 
$F_{\rm{ref}}$ increases together with the line contribution,
$F_{\rm{line}}$, when the primary source height decreases, and 
eventually dominates.

In Figure~\ref{fig_30degree}, EWs versus three different definitions of the
continuum are shown for the sake of illustration:
thick solid lines refer to EWs with respect to the total
underlying continuum, taking into account the solid angle distortion due
to the spacetime curvature; thin solid lines have been computed in a
similar manner, but with respect to the direct continuum only; finally,
the dash-dotted lines have been plotted with respect to the direct
continuum only, and they do not include the solid angle distortion
effect. It is worth noticing that, due to efficient emission from
the innermost region, in the extreme Kerr case one obtains EW values
very much enhanced at low $h$'s.

Table~\ref{tab_1} reports integral quantities (centroid energy, line
width, and equivalent width with respect to the total continuum) of the
line profiles for selected values of the parameters $h$ and
$\theta_{\rm{o}}$. The table refers to a disc extending up to $100$
gravitational radii and a maximally spinning black hole; values for a
static hole (with $r_{\rm{in}}=6m$) are given in parentheses for
comparison. In Table~\ref{tab_1}, $E_{\rm{cen}}$ and $\sigma$ are expressed
in keV while EW is in eV.

Finally, a word of caution is needed here on the effect of the iron
abundance. The EW strongly depends on abundance (e.g.\ Matt, Fabian
\& Reynolds 1997) but fortunately the other integral parameters of the
line do not so much. Moreover, the reflection component depends on the
iron abundance in an easily recognizable way, i.e.\ changing the depth of
the iron edge (Reynolds, Fabian \& Inoue 1995). Strong constraints have
been derived in the case of MCG--6-30-15 (Lee et al., 1999). Therefore one
can hope to separate the influence of poorly known iron abundances from
other effects.

\section{Conclusions}

We calculated the relativistic effects on both the iron line and the
reflection continuum emitted in the innermost regions of accretion discs
around spinning black holes in AGN and BHCs. The calculations are fully
relativistic with respect both to the primary emission (illumination
from a central source) and to the secondary one (disc reflection), so
that the line profiles are in this sense computed self-consistently. 

We found that the adopted geometry of the source copes very well
with observed widths and energy centroids of the spectral features around
6.4 keV. However, the final assessment of one of the few
viable models requires more detailed comparisons than it has been possible
so far with the currently available data. Predicted values should be
compared against high-resolution data, together with the results of
alternative scenarios, such as quasi-spherical accretion and further 
line broadening due to Comptonization. 

The development of a {\sc XSPEC} compatible code, making use of a big
atlas of
geodesics, is in progress. This will enable a fast fitting of the data.

The results presented here are thus relevant for the near future
high-sensitivity X-ray observatories, like {\it XMM}, as far as the iron
line is concerned. We have to wait for missions like {\it
Constellation-X\/}, with its very large sensitivity and broad band
coverage, in order to simultaneously examine both the iron line and the
reflection continuum in the desired detail.

\section*{Acknowledgments}
\def\lb#1{{\protect\linebreak[#1]}}
V\,K acknowledges support from grants GACR 205/\lb{2}97/\lb{2}1165 and
202/\lb{2}99/\lb{2}0261 in Prague.

\section*{APPENDIX: Fitting formulae for the emissivity laws and
integral quantities}

Local emissivities of the disc surface have been plotted in
Fig.~\ref{fig_emis}. In Table~\ref{tab_2} we provide a practical fit of
the emissivities, which can be useful in calculations. We used a simple
law of the type:
\begin{equation}
\epsilon(r)=c_1r^{-\lambda_1}+c_2r^{-\lambda_2}
\label{eq_emis}
\end{equation}

The adoption of function~(\ref{eq_emis}) for the fitting enables
comparisons with the power laws, which are commonly used in standard line
profile calculations. Attempts to derive the emissivity dependence from
radius by ``inverting'' observed line profiles have also been made, e.g., by
\v{C}ade\v{z} et al. (1999), Dabrowski et al. (1997), and Mannucci,
Salvati \& Stanga (1992).

%-----------------------------
\begin{table*}
\centering
\label{tab_2}
%\vspace{0.5pt}
\begin{tabular}{lccccccccccc}
\hline
\hline
   ~   &  $c_1$  &  $\lambda_1$  &  $c_2$  &  $\lambda_2$  &  \\
\hline
$h = 2m$  & $0.02268$ & $38.38706$    & $0.000122$    & $6.14786$ & \\
$h = 3m$  & $0.00146$ & $24.47516$    & $0.000059$    & $0.00006$ & \\
$h = 4m$  & $0.00324$ & $30.85495$    & $0.000039$    & $4.70109$ & \\
$h = 5m$  & $0.00640$ & $35.36543$    & $0.000025$    & $4.49187$ & \\
$h = 6m$  & $0.00968$ & $34.27642$    & $0.000013$    & $3.75576$ & \\
$h = 8m$  & $0.27139$x$10^{-4}$ & $8.74175$ & $0.01662$x$10^{-4}$ &
$1.74206$ & \\
$h = 10m$ & $0.99815$x$10^{-5}$ & $6.05660$ & $0.04793$x$10^{-5}$ &
$1.14952$ & \\
$h = 12m$ & $0.42940$x$10^{-5}$ & $4.42939$ & $0.01590$x$10^{-5}$ &
$0.81154$ & \\
$h = 15m$ & $0.27182$x$10^{-5}$ & $4.64923$ & $0.01451$x$10^{-5}$ &
$0.89655$ & \\
$h = 20m$ & $0.98182$x$10^{-6}$ & $3.66965$ & $0.05796$x$10^{-6}$ &
$0.72744$ & \\
$h = 100m$& $0.59002$x$10^{-5}$ & $1.05533$ &$-0.59023$x$10^{-5}$ &
$1.05622$ & \\
$h_{\rm cl} = 5m$ 
          & $0.10084$x$10^{-4}$ & $0.29168$ &$-0.09948$x$10^{-4}$ &
$0.28796$ & \\
\hline
\hline
\end{tabular}
\caption{Results of fitting the emissivity laws for different
source heights (see the Appendix for detailed explanation).
All coefficients have been obtained for a maximally rotating hole, except
for the row $h_{\rm cl}$=5m which refers to the purely classical (i.e.\
euclidean) case with the rays not distorted by gravitation.
For $r>6m$ (which is the case of stable orbits in Schwarzschild metric)
the emissivity is only weakly affected by the BH spin, while the effect
of rotation is substantial for small values of $h$.}
%\vspace(0.5pt)
\end{table*}

%-----------------------------
\begin{table*}
\centering
\label{tab_3}
%\vspace{0.5pt}     
\begin{tabular}{cccccccccccc}
\hline
\hline
\multicolumn{1}{c}{$\cos{\theta_{\rm{o}}}$} &
\multicolumn{1}{r}{ ~ } &
\multicolumn{1}{c}{ $\!|\!$ } &
\multicolumn{2}{c}{$a_2$} &
\multicolumn{1}{c}{ $\!|\!$ } &
\multicolumn{2}{c}{$a_1$} &
\multicolumn{1}{c}{ $\!|\!$ } &
\multicolumn{2}{c}{$a_0$}  \\
%\multicolumn{2}{c}{ ~ } & \\
\hline
& $E_{\rm{cen}}$ &&    0.00243  & (0.000594)  &  
                  &   $-$0.112  & ($-$0.0301) & & 8.05 & (6.88)\\
0.1 &   $\sigma$ &&  $-$0.00103 & ($-$8.77x$10^{-5}$) & 
                  & $-$0.000877 & ($-$0.019)  &  & 1.74 & (1.37)\\
& EW             &&     2.85    & ($-$0.0979) &  
                  &     $-$88.8 & (2.68) &  & 687 & (21.4) \\
\hline
& $E_{\rm{cen}}$ &&   $-$0.0011 & (0.000198)  & 
                  &    0.0251 & ($-$0.0162) & & 6.43 & (6.71)  \\
0.3 &  $\sigma$  &&  0.000643 & (0.000111)    & 
                  &   $-$0.0509 & ($-$0.0217) & &  1.83 & (1.35) \\
&  EW            &&      2.07 & ($-$0.173)    & 
                  &     $-$65.1 & (4.74) &  &  555 & (42.6) \\
\hline
& $E_{\rm{cen}}$ &&  $-$0.00414 & (3.69x$10^{-5}$) & 
                  &     0.134 & ($-$0.00223) & &   5.27 & (6.37) \\
0.5 &  $\sigma$  &&   0.00104 & (0.000136)         & 
                  &   $-$0.0598 & ($-$0.0203) & &  1.73 & (1.21) \\
&  EW            &&       1.5 & ($-$0.206)         &
                  &  $-$47.4 & (5.68) & & 445 & (53.6) \\
\hline
& $E_{\rm{cen}}$ &&  $-$0.00614 & ($-$0.000161)    & 
                  &     0.208 & (0.014) & &  4.38 & (5.95) \\
0.7 &  $\sigma$  &&   0.00118 & (0.000111)         & 
                  &   $-$0.0616 & ($-$0.0167) & &  1.58 & (0.999) \\
&  EW            &&      1.21 & ($-$0.224)         & 
                  &     $-$38.3 &  (6.44) & & 384 & (55.8) \\
\hline
& $E_{\rm{cen}}$ &&  $-$0.00769 & ($-$0.00042)     & 
                  &  0.266 & (0.031) & & 3.62 & (5.52) \\
0.9 &  $\sigma$  &&   0.00127 & (1.34x$10^{-19}$)  & 
                  &  $-$0.0647 & ($-$0.01) & &  1.44 & (0.71) \\
&  EW            &&     0.952 & ($-$0.239)         & 
                  & $-$30 & (6.98) & &  327 & (56) \\
\hline
\hline
\end{tabular}
\caption{Coefficients of the least-square polynomial fits to
$h$-dependences of $E_{\rm{cen}}$, $\sigma$ and EW, as predicted
in our model. Angular-momentum parameter is $a=0.9981m$; for comparison,
the values resulting from computations with Schwarzschild metric are
given in parentheses. }
\end{table*}

In Table~\ref{tab_3} we provide coefficients of the least-square fitting
for the observable quantities: $E_{\rm{cen}}(h)$, $\sigma(h)$, and
EW$(h)$. Here we used quadratic polynomials of the form
\begin{equation}
a_0+a_1\,h/m+a_2\,(h/m)^2,
\end{equation}
which can approximate $h$-dependences in the
range $4m\leq{h}\leq20m$ with sufficient accuracy. For the
whole interval, up to $h=100m$, we used more precise spline fits; the
corresponding Matlab script is available from the authors (this can be
used for the numerical inversion to obtain parameters of the model,
i.e.\ $a/m$, $\theta_{\rm{o}}$, and $h/m$, in terms of the three
observables).

\end{document}